\documentclass[a4paper]{article}
\usepackage{graphicx}
\usepackage{onecolpceurws}

\title{Cloud--based Fault Detection and Classification for Oil \& Gas Industry}

\author{
  Athar Khodabakhsh \\ 
  Computer Science Dept. \\
  Ozyegin University, Istanbul, Turkey \\
  athar.khodabakhsh@ozu.edu.tr
\and
  Ismail Ari  \\ 
  Computer Science Dept. \\
  Ozyegin University, Istanbul, Turkey \\
  ismail.ari@ozyegin.edu.tr
\and
  Mustafa Bakir \\ 
  Process Improvement and Software Dept. \\
  TUPRAS, Kocaeli, Turkey\\ 
  mustafa.bakir@tupras.com.tr
}
\institution{}

\begin{document}
\maketitle

\begin{abstract}

Oil \& Gas industry relies on automated, mission--critical equipment and complex systems built upon their interaction and cooperation. To assure continuous operation and avoid any supervision, architects embed Distributed Control Systems (DCS), \emph{a.k.a.} Supervisory Control and Data Acquisition (SCADA) systems, on top of their equipment to generate data, monitor state and make critical online \& offline decisions. 

In this paper, we propose a new Lambda architecture for oil \& gas industry for unified data and analytical processing on data received from DCS, discuss cloud integration issues and share our experiences with the implementation of sensor fault--detection and classification modules inside the proposed architecture. 

\end{abstract}
\vskip 32pt

\section{Introduction}

Industry 4.0 revolution manifests that Internet of Things (IoT), Big Data, and Cloud technologies should be used together to deliver a digital transformation and establish smarter operation for all sectors including oil \& gas. However, data analysis for petrochemical industry contains \emph{all} of the big data challenges (4V's: volume, velocity, variety, and veracity) that require ability to scale, integrate and reconcile data in real--time. Fortunately, the industry is ready to make this transformation~\cite{Schwartz-opportunity}, but it is looking for quick and effective ways to complete the task. 

Due to mission--criticality of their processes, oil \& gas businesses have already implanted thousands of sensors inside and around their physical systems~\cite{Perrons-Data}. Raw sensor data continuously streams via DCS and SCADA systems measuring temperature, pressure, flow rate, vibration, and depth \emph{etc.} of drills, turbines, boilers, pumps, compressors, and injectors. A right quantity of data should be extracted, transformed and passed from data quality tests before loading (\emph{i.e.} ETL process). To achieve success in these steps, one also needs to understand the nature of processes, devices and data to support real--time decision making~\cite{Feblowits}. 

Algorithmic designs also need to change to support incremental updates over streams. Within the last decade, special ''stream mining'' versions~\cite{aggarwal} of rule--mining~\cite{Ari-stream}, pattern recognition, classification and clustering algorithms have been developed and embedded into emerging data architectures. 

In this paper, we focus on the veracity, \emph{i.e.} correctness, property of oil \& gas big data. For this purpose data has to be cleaned and reconciled before it can be used to train stream mining algorithms to generate real--time models and results.  

\section{Big Data Platform Architecture for Oil \& Gas Analytics}

To manage data storage, processing, and analytics at scale oil \& gas industry recently started experimenting with open--source distributed frameworks such as Apache Hadoop~\cite{hadoop-org}, Spark~\cite{spark} and Ignite~\cite{ignite}. Apache Hadoop  consists of Hadoop Distributed File System (HDFS), MapReduce, HBase, Hive, and other system management modules. Apache Spark solves online processing problems of disk--based Hadoop by implementing MapReduce layer entirely in--memory, but its Resilient Distributed Datasets (RDDs) are immutable. Apache Ignite provides shared, mutable in--memory RDDs called data grid, which is a NoSQL key--value store (partitioned hashmap) that can be employed for providing distributed caching or complementing Apache Spark. While these technical developments in distributed systems and their ease of availability is useful for the oil \& gas sector, there are tens of open--source data processing \& mining projects with no unified view about how to leverage them coherently, because:

\begin{itemize}

\item oil \& gas industry has not developed a comprehensive data architecture to complete its digital transformation neither in the drilling nor refining businesses, and,
\item the integration with private and public clouds is not discussed clearly.
\end{itemize}

We propose to use Lambda architecture as the unified data and analytics architecture for oil \& gas, which consists of three layers: \emph{batch processing layer} for offline data, \emph{serving layer} for preparing indexes and views, and \emph{speed layer} for real--time processing~\cite{Kiran-processing}. We also introduce Lambda architecture implemention alternatives in private and public clouds.

\begin{figure}[ht]
\begin{center}
\includegraphics[height=6.5cm]{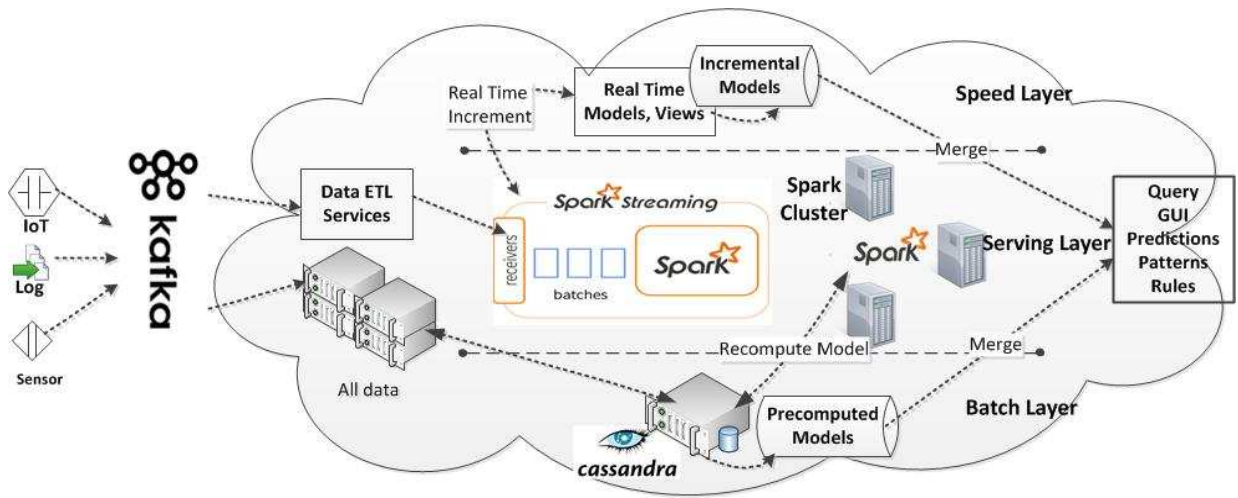}
\caption{Proposed Lambda architecture for oil \& gas data analysis.}
\label{fig-architecture}
\end{center}
\end{figure}

The proposed architecture is depicted in Figure~\ref{fig-architecture} for a private cloud. Apache Kafka~\cite{kafka} is an open--source publish--subscribe engine for building real--time data pipeline and stream processing applications. Sources of data streams such as sensors, log files, and IoT act as the publishers.  
Applications in the speed and batch layers subscribe to Kafka for pulling the data and processing them in real--time. Spark Streaming~\cite{spark} receives live data streams and divides them into smaller batches. This data is also stored in a distributed database management system such as Apache Cassandra~\cite{cassandra}, which replicates it internally to several nodes to ensure reliability. Distributed algorithms can now have access to replicas via MapReduce framework of Apache Spark API. Complex analytics such as pattern recognition, classification, rule extraction, and fault detection are performed in serving layer. MlLib~\cite{mllib} is an open--source machine learning library which works on top of Apache Spark as well as Cassandra. The resulting real--time (online) models in speed layer and precomputed (offline) models in batch layer are merged for visualization and prediction purposes.

The private cloud data system can be integrated with public cloud services such as Amazon Web Services (AWS)~\cite{amazon}. Each open--source project shown in Figure~\ref{fig-architecture} has a public cloud version: AWS--Kinesis is cloud version of Kafka, AWS--EMR (Elastic MapReduce) is data processing framework for MapReduce, AWS--S3 (Simple Storage Service) is for storing batches and log files similar to Cassandra.

\section{Fault Detection and Classification Modules}

Measurement errors in sensor data are categorized into two main classes: random errors (statistical noise) and gross errors (due to problems in sensor devices)~\cite{Narasimhan-Data}. When a sensor starts malfunctioning, its values must either be discarded or reconciled until the sensor is readjusted or replaced. 

\begin{figure}[ht]
\begin{center}
\includegraphics[height=8cm,keepaspectratio]{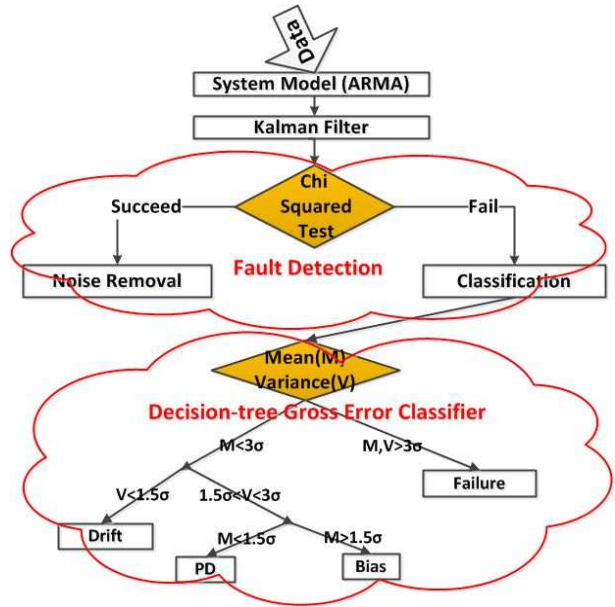}
\caption{Block diagram for fault detection and error classification. }
\label{fig-blockDiagram}
\end{center}
\end{figure}

We have used statistical pattern recognition techniques including  Least Square Estimation (LSE), Auto--Regressive Moving Average (ARMA), and Kalman Filters (KF) to extract system properties and develop fault detection models over real and synthetically--generated refinery data. Locating which sensor makes what type of error improves overall system reliability for the refinery and improves predictive maintenance performance~\cite {Feblowits}.
Figure~\ref{fig-blockDiagram} shows the block diagram of our methodology for fault detection and error classification. For extracting a system's properties, ARMA model described by Equation \ref{eq:1} is used to obtain system's characteristics, where $k$ represents the time index,  $y_k$ is the output, and $x_k$ is the input at time $k$. The coefficients $\alpha_i$ and $\beta_j$ are extracted using LSE, where $i$ and $j$ represent system's memory.

Modeling the time--varying refinery plants, the anomalies can be detected with statistical hypothesis tests (chi--squared test) for fault detection and noise removal. If the hypothesis fails, then a gross error is detected.

\begin{equation}\label{eq:1}
‎y_k+ \alpha_1 y_{(k-1)}+\dots+\alpha_n y_{(k-n)}= \beta_0 x_k+\dots+ \beta_m x_{(k-m)}  
\end{equation}

\newblock

The plant is converted into a state--space representation using the properties extracted from the ARMA model. Kalman Filter (KF) is applied on top of this time--varying system to track its dynamic state. It is possible to utilize the \emph{innovation} properties of KF for fault detection. Innovations are normally distributed which are computed in the correction step of the KF and are subject to chi--squared test. The global test is defined by Equation \ref{eq:2}, where $V_k$ is \emph{innovation} covariance and $v_k$ is \emph{innovation} residual at time $k$. 

\begin{equation}\label{eq:2}
 \gamma=v_k^T V_k^{-1} v_k
\end{equation}

If $\gamma$ exceeds the criterion of corresponding probability, the hypothesis fails, gross error is detected, and the sensor location is identified.

Data with gross errors are then analyzed using a decision tree classifier to classify errors into four types called Bias, Drift, Precision Degradation (PD) and Failure. The variations need to be less than 3--$\sigma$ according to normal distribution. Decision--tree classifier works as follows:

\begin{itemize}
\item if mean variation is $<$3$\sigma$ and variance is $<$$\sigma$ gross error type is Drift, 
\item if mean is $<$1.5$\sigma$ and variance is  between 1.5$\sigma$ and 3$\sigma$ type is PD, 
\item if mean is $>$1.5$\sigma$ and variance is  between 1.5$\sigma$ and 3$\sigma$ it is classified as Bias, 
\item otherwise the sensor has Failed completely. 
\end{itemize}

\subsection{Model Updates}

\begin{figure}[ht]
\begin{center}
\includegraphics[height=5cm]{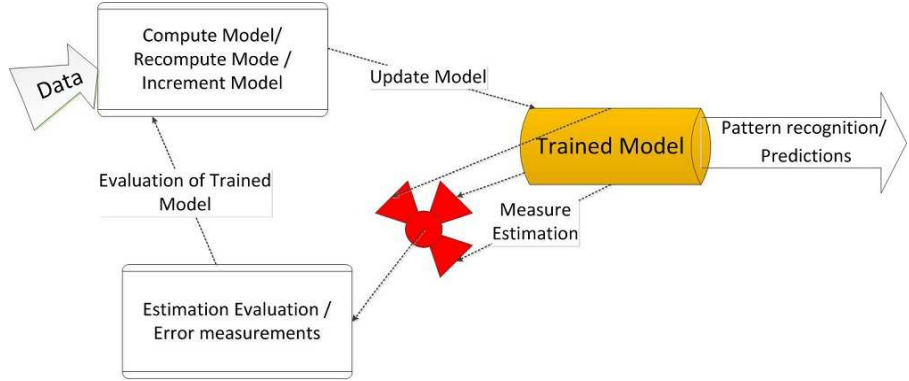}
\caption{Incremental training and evaluation model. }
\label{fig-training}
\end{center}
\end{figure}

Figure \ref{fig-training} details the incremental training and evaluation of a model as shown in Fig.\ref{fig-architecture}. The trained model is tested on data stream and estimation error is calculated~\cite{Tan-Mining}. According to evaluation result, the model may be passed to recomputing module, incrementing the current model or computing a new model based on current state of the system. 

\newblock

\subsection{Complexity of Algorithms}
Algorithms developed for data streams must process data under certain time and space restrictions. Three algorithms were used in our fault detection and classification approach. ARMA model is constructed from input data and the system's characteristics is then passed to KF for fault detection. If gross error is detected, then a classification algorithm is activated to classify the detected anomalies. We used a decision tree algorithm, whose worst--case complexity is $\Omega({log n})$ similar to binary trees, where $n$ is the number of items in the tree. KF algorithm for tracking system dynamicity and fault detection is $\Omega({n^3})$~\cite{Vai-Cmpx} in worst case, where $n$ is the matrix dimension. Since we used the packaged JBLAS matrix multiplication and this package is an optimized library for Java, the complexity is close to $\Omega({mn^2})$.

\subsection{Preliminary Results}

\begin{figure}[ht]
\begin{center}
\includegraphics[height=6cm,width=15cm,keepaspectratio]{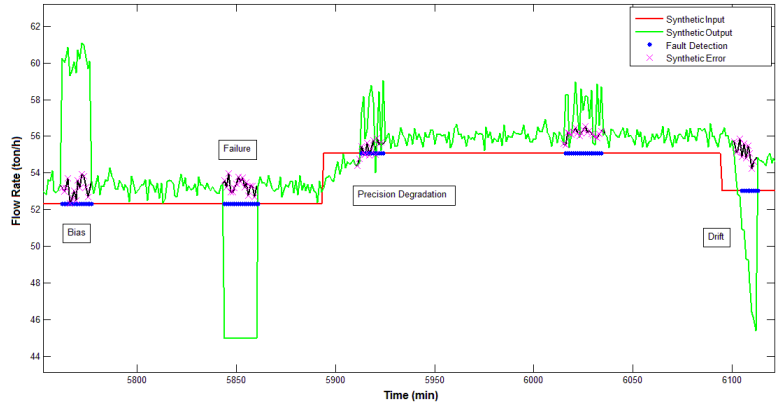}
\caption{Gross error detection (GED) and classification using synthetic refinery data.}
\label{fig-synthetic}
\end{center}
\end{figure}

\begin{figure}[ht]
\begin{center}
\includegraphics[height=5cm,width=15cm,keepaspectratio]{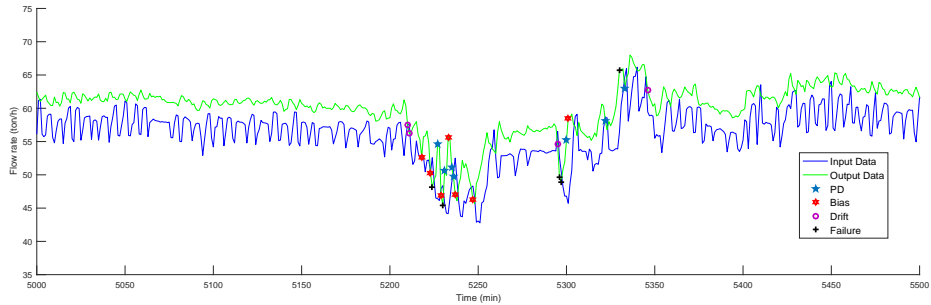}
\caption{Gross error detection and classification on real data.}
\label{fig-errorDetection}
\end{center}
\end{figure}

We implemented these modules on top of a Complex Event Processing (CEP) engine, called ESPER~\cite{ESPER} and tested over data streams. Figure~\ref{fig-synthetic} shows our results for the synthetic data extracted from characteristics of real data. Synthetically generated gross error events were  inserted at random points to evaluate accuracy of detection and classification. Then our model trained on synthetic data was validated with real data.  Figure~\ref{fig-errorDetection} shows fault detection and classification results on top of real data. Both figures illustrate that we can detect gross errors and classify them correctly. We reserve the details about the accuracy of our methods as future work.

\newblock

\newblock

\section{Related Work}

The fault detection and classification algorithms in our approach, integrate prior information of system to plant model using statistical pattern recognition and detect anomalies in the system. Alenany, et. al~\cite{Ale-lse} propose an efficient subspace identification method, but do not integrate anomaly detection and apply it to data streams. A maximum likelihood method is proposed in~\cite{Lar-likelhd} for obtaining accurate statistical model of system. In~\cite{Zha-dyn}, a PCA-based error detection is used that improves error detection on different modes of operations and improves the constructed model. In this work, we used ARMA modeling and KF tracking for system identification and gross error detection online, over streaming sensor data. We also proposed a Lambda architecture for unified stream data processing and analytics for oil \& gas industry. Another related work for building oil \& gas ontologies~\cite{Norwegian} and addressing data integration issues includes POSC Caesar Association (PCA)'s ISO 15926 standardization efforts. 

\section{Conclusion}

Oil \& Gas industry demands effective methods to complete its Industry 4.0 digital transformation. This leads them to resort to open--source big data tools. In this paper, we proposed an architecture to solve big data problems seen in oil refinery sensor data and implemented the architecture in private cloud utilizing open--source software. The results are applicable to oil drilling industry as well. We developed real--time analytical models for detecting and classifying gross errors. Our contributions are enablers of other stream mining algorithms and provisioning of ``Refinery as a Service''. In the future, we plan to continue and complement our current work with other machine learning and deep learning algorithms.

\section{Acknowledgements}

This work in progress is funded and supported by TUPRAS Inc. Oil Refinery in Turkey. We would like to thank process and software experts Burak Aydogan and Mehmet Aydin from TUPRAS for providing us the DCS data used in this research and also Dr. Ali Ozer Ercan for giving valuable feedbacks. 


\end{document}